\documentclass{emulateapj}
\input{psfig.sty}
\usepackage{natbib}
\newcommand{\kms}{\,{\rm km\,s^{-1}}}

\newcommand{\beq}{\begin{equation}}
\newcommand{\eeq}{\end{equation}}
\newcommand{\ba}{\begin{eqnarray}}
\newcommand{\ea}{\end{eqnarray}}
\def\spose#1{\hbox to 0pt{#1\hss}}
\newcommand{\lta}{\mathrel{\spose{\lower 3pt\hbox{$\mathchar"218$}}
      \raise 2.0pt\hbox{$\mathchar"13C$}}}
\newcommand{\gta}{\mathrel{\spose{\lower 3pt\hbox{$\mathchar"218$}}
      \raise 2.0pt\hbox{$\mathchar"13E$}}}
\def\simlt{\mathrel{\rlap{\lower 3pt\hbox{$\sim$}}\raise 2.0pt\hbox{$<$}}}
\def\simgt{\mathrel{\rlap{\lower 3pt\hbox{$\sim$}} \raise 2.0pt\hbox{$>$}}}
\lefthead{Marta Volonteri}
\righthead{Signatures of gravitational recoil}
\makeatletter

\newenvironment{figurehere}
  {\def\@captype{figure}}
  {}
\makeatother
\begin{document}
\title{Gravitational recoil: signatures on the massive black hole population}
\author{Marta Volonteri\altaffilmark{1}}
\altaffiltext{1}{University of Michigan, Astronomy Department, Ann Arbor, MI, 48109}

\begin{abstract}
In the last stages of a black hole merger, the binary can experience a recoil due to asymmetric emission of gravitational radiation. Recent numerical relativity simulations suggest that the recoil velocity can be as high as a few thousands kilometers per second for particular configurations. We consider here the effect of a {\it worst case scenario} for orbital and phase configurations on the hierarchical evolution of the massive black hole (MBH) population. The orbital configuration and spin orientation in the plane is chosen to be the one yielding the highest possible kick.  Masses and spin magnitudes are instead derived self-consistently from the MBH evolutionary models. If seeds form early, e.g. as remnants of the first stars, almost the totality of the first few generation of binaries are ejected. The fraction of \emph{lost} binaries decreases at later times due to a combination of the binary mass ratio distribution becoming shallower, and the deepening of the hosts potential wells. If seeds form at later times, in more massive halos, the retention rate is much higher. We show that the gravitational recoil does not pose a threat to the evolution of the MBH population that we observe locally in either case, although high mass seeds seem to be favored.  The gravitational recoil is instead a real hazard for (i) MBHs in biased halos at high-redshift,  where mergers are more common, and the potential wells still relatively shallow. Similarly, it is very challenging to retain (ii) MBHs merging in star clusters.
\end{abstract}


\section{Introduction}
Today, MBHs are ubiquitous in the  nuclei of nearby galaxies \citep[see, e.g.,][]{richstone1998}. If MBHs were also common in the past, as implied by the large population of quasars, and if their host galaxies experience multiple mergers during their lifetime, as expected in the  currently favored cold dark matter  hierarchical cosmologies, then MBH mergers are expected to be relatively common during the cosmic history.  

The results of recent numerical relativity simulations question whether MBH mergers represent a threat to the evolution of the MBH population that we observe today.  In the final phase of a black hole merger, the inspiral is driven by emission of gravitational radiation. Gravitational waves carry, in general,  a non-zero net linear momentum, which establishes a preferential direction for the propagation of the waves. As a consequence, the center of mass of the binary recoils in the opposite direction \citep{redmountrees}, possibly causing the ejection of MBHs from the potential wells of their host galaxies  \citep[e.g.,][]{Madauetal2004,MadauQuataert2004, Merrittetal2004, VolonteriRees2006, Haiman2004, sb2007}.   

The calculations of the magnitude of the recoil have been oscillating in literature by a few orders of magnitude from the early calculations in the Newtonian regime \citep{Fitchett1983}, to the latest analytical approaches \citep{Favataetal2004,Blanchetetal2005,Gopu2006}. The advent of numerical relativity is now leading to a convergence in the estimates of the recoil. 
Schwarzschild, i.e., non-spinning, black holes \citep[e.g.,][]{Bakeretal2006} are expected to recoil with velocities below 200 $\rm{km\,s^{-1}}$, and a similar range is expected for black holes with low spins, or with spins (anti-)aligned with the orbital axis (Figure \ref{fig01}). However, when the spin vectors have opposite directions and are in the orbital plane, the recoil velocity can be as large as a few thousands $\rm{km\,s^{-1}}$ \citep{campanelli2007b,gonzalez2007,campanelli2007}.

In this paper we assess {\it if} standard models for MBH evolution (mass and spin) can accommodate a high recoil, and {\it when} such high recoils are indeed creating hazards during the hierarchical growth of the MBH population.  We trace the mass and spin evolution of MBHs from early times, within a plausible scenario for the hierarchical assembly, growth, and dynamics of MBHs in a $\Lambda$CDM cosmology, that has been shown to capture many features of the MBH population (e.g., luminosity function of quasars, MBH mass density, core formation due to MBH binary mergers). The main features of the models have been discussed in \cite{VHM,Volonterietal2005, VolonteriRees2006, gw3}, and references therein.  During a MBH merger, we assume the orbital and phase\footnote{\cite{campanelli2007b} also pointed out that the recoil velocity is not uniquely determined solely by the magnitudes and relative directions of the black hole spins in the orbital plane. The recoil was found to vary sinusoidally with the angle between the orbital plane (parallel to the black hole spins) and the initial linear momenta of the holes.} configuration that leads to the largest recoil magnitude. Black hole masses and spins magnitudes are instead self-consistently determined by the merger and accretion history of the MBHs. 

We compare here  two different models of MBH formation, one that predicts MBHs to form early, and with low mass, as remnants of the first generation of stars (PopIII stars, cfr. model VHM in Sesana et al. 2007), one which predicts MBHs to form later and with larger masses (cfr. model BVRlf in Sesana et al. 2007), in halos with low angular momentum, prone to large scale instabilities \citep{BegelmanVolonteriRees2006}.  We refer the reader to  \cite{gw3} for an exhaustive description of the models (including accretion properties and dynamical evolution). We have assessed models with MBH formation more/less efficient than the reference cases. In case of very efficient MBH formation, the mass density on MBHs at $z\simeq3$ is much larger than expected from the observed quasar evolution \citep{YuTremaine2002, Merloni2004}. In models where MBHs form in very small numbers \citep[e.g., BVRhf in][]{gw3}, mergers are very rare events, and happen at later times, with small mass ratios. Clearly, the gravitational rocket is much less of a hazard in models which start from fewer seeds at early times.  
\begin{figurehere}
\centerline{
\psfig{file=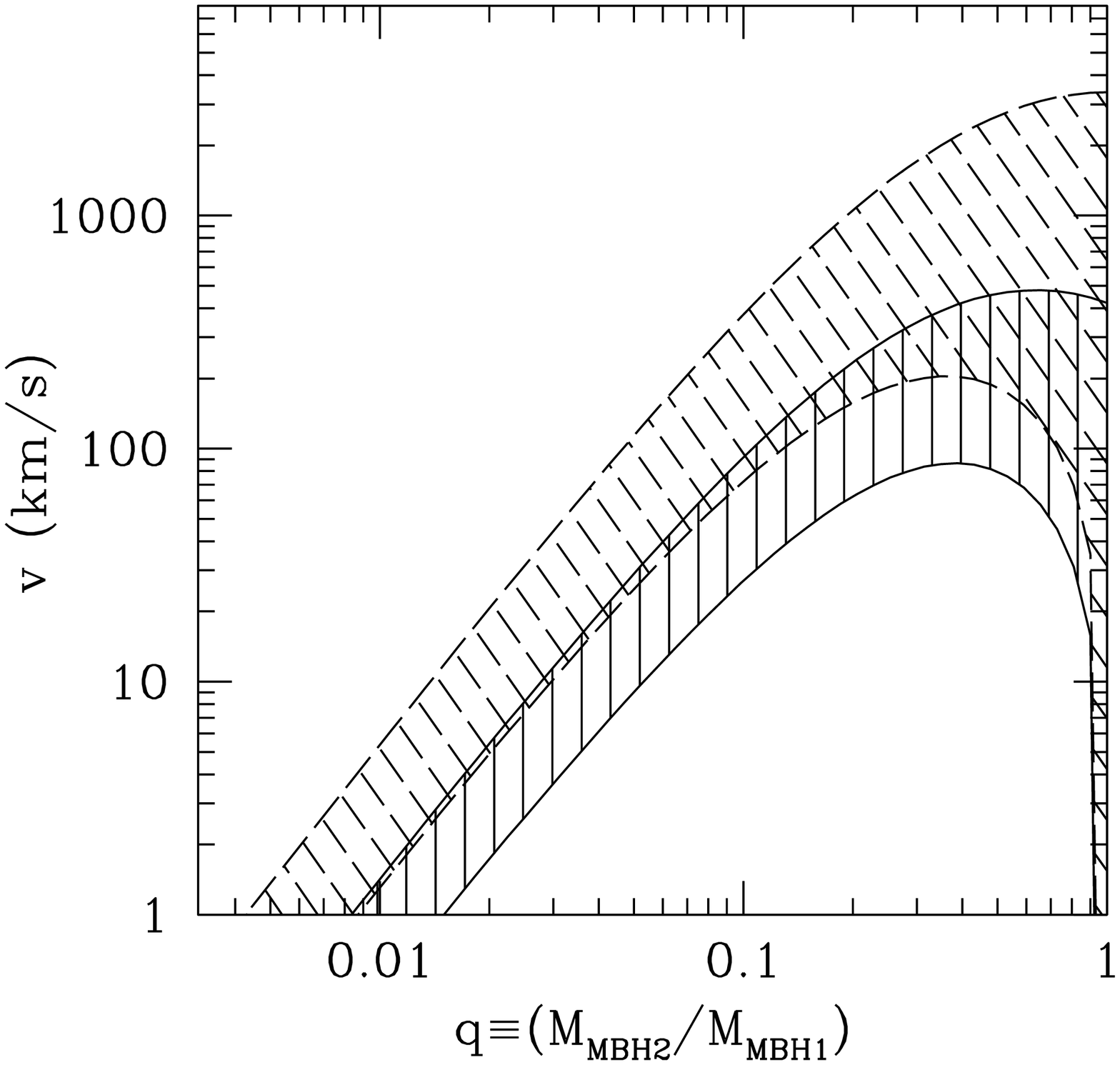,width=2.8in,height=2.8in}}
\caption{\footnotesize Recoil velocity of spinning black holes as a function of binary mass ratio, $q$. {\it Solid curves:} spin axis aligned (or anti-aligned) with the orbital angular momentum (Baker et al. 2007). {\it Dashed curves:} spin axis in the orbital plane (Campanelli et al. 2007).  For every mass ratio we plot the combination of spins, $\hat a_1$ and $\hat a_2$ which minimizes ({\it lower curves}) or maximizes ({\it upper curves}) the recoil velocity.}
\label{fig01}
\end{figurehere}

\section{Merging MBH mass ratios and spins}
During a galactic merger, dynamical friction drags in the satellite, along with its central MBH, towards the center of the more massive progenitor. The orbital evolution of the galaxy pair leads also to the destabilization of gas in the merging systems. In simulations, this gas is seen to accumulate towards the centers of the system and fuel the MBHs in both galaxies \citep[e.g.,][]{DiMatteo2005, steliosetal2005, Mayeretal2006}.  When the orbital decay is efficient,  the satellite hole moves towards the center of the more massive progenitor, leading to the formation of a bound MBH binary. 

The efficiency of dynamical friction decreases with the mass ratio of the merging galaxies, so nearly equal mass galaxy mergers (``major mergers'') lead to efficient binary formation within short timescales (i.e., shorter than the Hubble time),  while in ``minor mergers''  it can take longer than an Hubble time to drag  the  satellite hole to the center.  This effects must be convolved with the mass ratio probability distribution. As the mass function of halos (and galaxies) is steep, the probability of halo mergers decreases with increasing mass ratio, that is, dynamically efficient major mergers are rare, while minor mergers, inefficient in forming MBH binaries, are common.

The combination of those effects leads to a distribution of mass ratios for merging black holes, which is at most redshifts dominated by systems with $q\equiv M_{MBH2}/M_{MBH1}\le 1$ in the decade around $q=0.1$. 
At early times, equal mass MBH mergers are more common, as most mergers involve MBHs which had no time to accrete  much mass yet. 
At later times binaries with small mass ratios are instead much more common, as some MBHs that started their orbital decay during high redshift minor mergers  (with an orbital decay timescale comparable to the Hubble time) finally find their way to the central MBH. The lower redshift mass ratio distributions are therefore much shallower \citep[cfr.][]{gw3}: high-speed recoils are common at high redshift, but the typical strength of the recoil decreases with time. Note that the recoil estimates by \cite{baker2007} and \cite{campanelli2007b, campanelli2007} both imply velocities below a few $\rm{km\,s^{-1}}$ for mass ratios $q<0.01$. 

The other two variables determining the recoil velocity (as we have fixed the relative inclination of the spins with respect to each other and the orbital plane) are the MBHs spin magnitudes\footnote{We express the spin through the dimensionless parameter $\hat a$, $S=\hat aGm^2_{\rm BH}/c$, $0\le \hat a\le 1$}.  MBH spins, in their cosmic evolution, are determined by the sequence of MBH mergers and accretion episodes that build up the MBH mass.  In general, mergers of similar mass  MBHs increase the spin, while the capture of smaller companions in randomly-oriented orbits tends to spin holes down \citep{HB2003}. Given our distribution of MBH binary mass ratios, binary coalescences alone do not lead to a dominant spin-up or spin-down of MBHs \citep[cfr.][]{Volonterietal2005}.  Gas accretion, on the other hand, seems to be more efficient in determining the distribution of MBH spins, due to the alignment between spin and angular momentum of the accretion disc induced by the Lense-Thirring precession.

In our simulations,  we assume that in an accretion episode the MBH spin is initially misaligned with respect to the angular momentum of the disc by a random angle, and that the alignment is relatively inefficient, following the scheme described in \cite{Volonterietal2005} (eq. 17-19; see Volonteri, Sikora \& Lasota 2007 for a more comprehensive model). The spin distributions that we derive are dominated by Kerr MBHs\footnote{For simplicity, we will assume in the following that a MBH may be spun up to a maximum equilibrium value of $\hat a=0.998$ (Thorne 1974), although we note that relativistic magnetohydrodynamic simulations for a series of thick accretion disc models  converge at a spin equilibrium of $\hat a\approx 0.93$ (Gammie, Shapiro, \& McKinney 2004). }, and are steeper for the spin of the primary (i.e. the most massive hole in the binary), as it typically grows by a larger factor during a given merger.

\section{Discussion}
\subsection{Signatures on the local MBHs population}
Our set of simulations allows to assess the influence of the gravitational recoil under realistic, although rather pessimistic, assumptions.  The fraction of merging MBHs that are actually ejected from their hosts, i.e., that have a recoil velocity larger than the escape speed from the host, has a strong dependence on redshift and on the MBH seeds formation mechanism (Figure \ref{fig1}). The escape velocity calculation assumes an NFW halo for the DM component. The baryonic component is modeled as an isothermal sphere truncated at the radius of the MBH sphere of influence, that is $r_{inf}=G M_{\rm MBH}/\sigma_g^2$, where $M_{\rm MBH}$ is the MBH mass and $\sigma_g$ is the velocity dispersion.

If seeds form early, in small proto-galaxies, almost the totality of the first few generation of binaries are ejected into the intergalactic medium. The fraction of ``lost'' binaries decreases at later times due to a combination of (i) the mass ratio distribution becoming shallower, and, (ii) the hierarchical growth of the hosts.  The ejected fraction decreases below 50\% at $z\simeq5$.
If seeds form at later times, in more massive halos, the retention rate is much higher from the beginning, and only about 40-50\% of binaries are ejected at all redshifts higher than $\simeq$3. 

\begin{figurehere}
\centerline{
\psfig{file=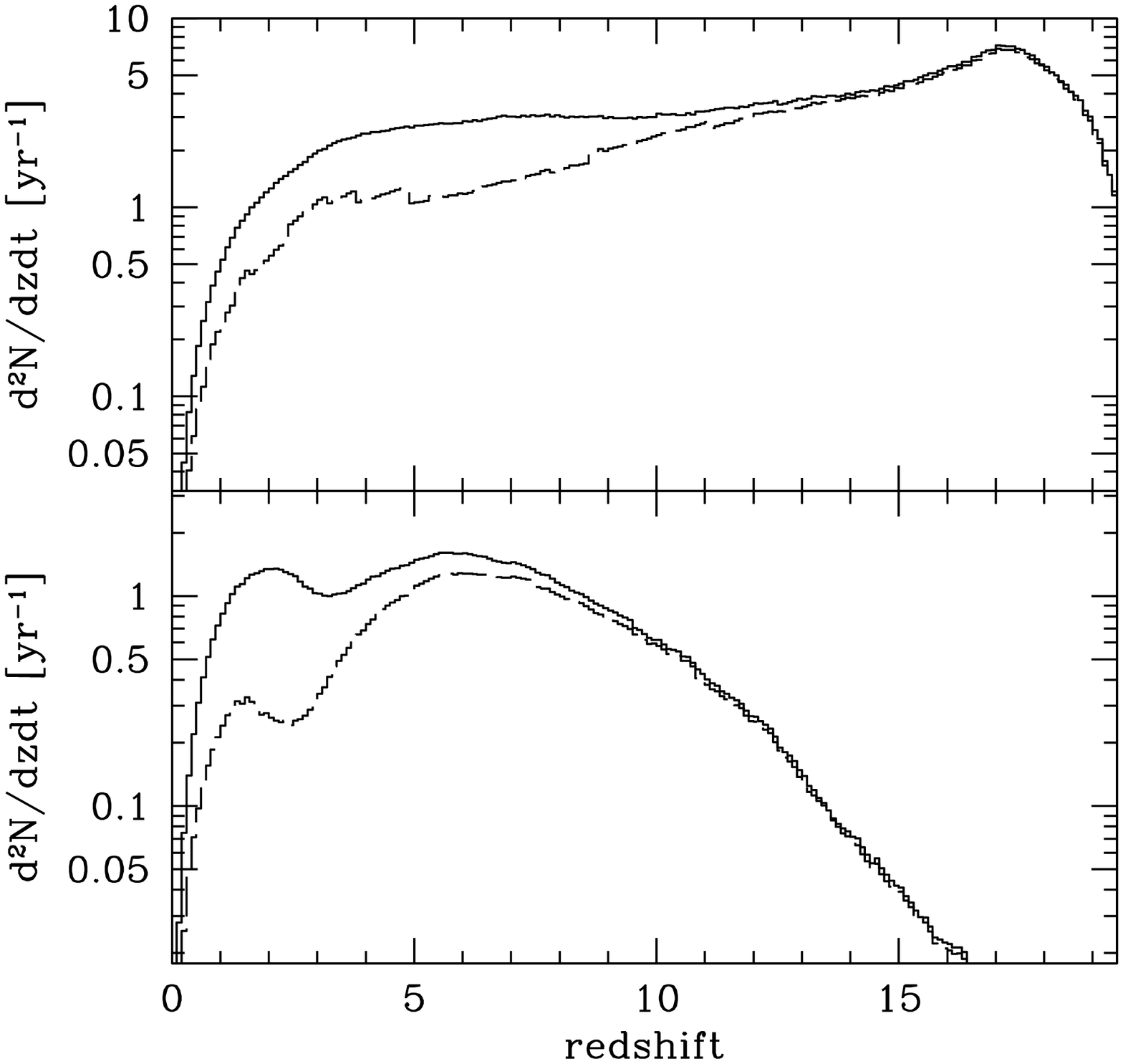,width=\columnwidth}}
\caption{\footnotesize Merger rate of binaries ({\it solid histogram}) and ejection rate of these binaries ({\it dashed histogram}), a binary is defined ejected if the recoil velocity is larger than the escape velocity frm the host. {\it Upper panel}: popIII remnants; {\it lower panel:} direct collapse.}
\label{fig1}
\end{figurehere}

Let us now quantify if these ejection rates indeed disturb the global evolution of the MBH population. The number of MBHs in binaries is a very small fraction of the population at most times (Figure \ref{fig2}) in all the considered cases. In fact, only about 40\% of MBHs ever experience a merger, and MBHs evolve in isolation for a large part of their lifetime. As, especially at high redshift, binaries represent the exception, rather than the rule, the possible ejection of most binaries before $z\simeq 5$ is not a threat to the evolution of the MBH population that has been detected in nearby galaxies.  Provided that one or few MBHs per galaxy merger tree \citep{menou2001} survive until $z\simeq 5$,  Soltan-type arguments,  which compare the mass density of MBHs in today quiescent galaxies with the mass density inferred from luminous quasars, imply that accretion alone can grow the MBH up to the supermassive variety (Yu \& Tremaine 2003).

Figure \ref{fig3} compares the simulated MBHs that are left in galaxies at $z=0$ to the current observational sample (samples from Ferrarese \& Ford 2005; and Tremaine et al. 2002, augmented for Lauer et al. 2006 additions; see Valluri et al. 2004 for uncertainties in the MBH mass measurements). The simulated sample broadly follows the same scaling with galaxy velocity dispersion, $\sigma_g$, of the observational sample. By visual inspection, the simulated MBHs display some ``outliers'' with masses below the predictions of the MBH mass - $\sigma_g$ scaling. Given the small observational sample, and the uncertainties on the MBH masses, we ran a Montecarlo test to compare the observational datasets to our simulations \citep[cfr.][]{stuart2006}; we conclude that statistically we cannot rule out that the simulated sample is compatible with observations (at the 2-$\sigma$ level), at least for MBHs hosted in galaxies with velocity dispersion larger than $\simgt100\kms$. Additionally, observations are biased against the detection of ``under-massive black holes'', with respect to the expectations of the MBH mass - $\sigma_g$ scaling, given the scaling of the radius of the sphere of influence. 
\begin{figurehere}
\centerline{
\psfig{file=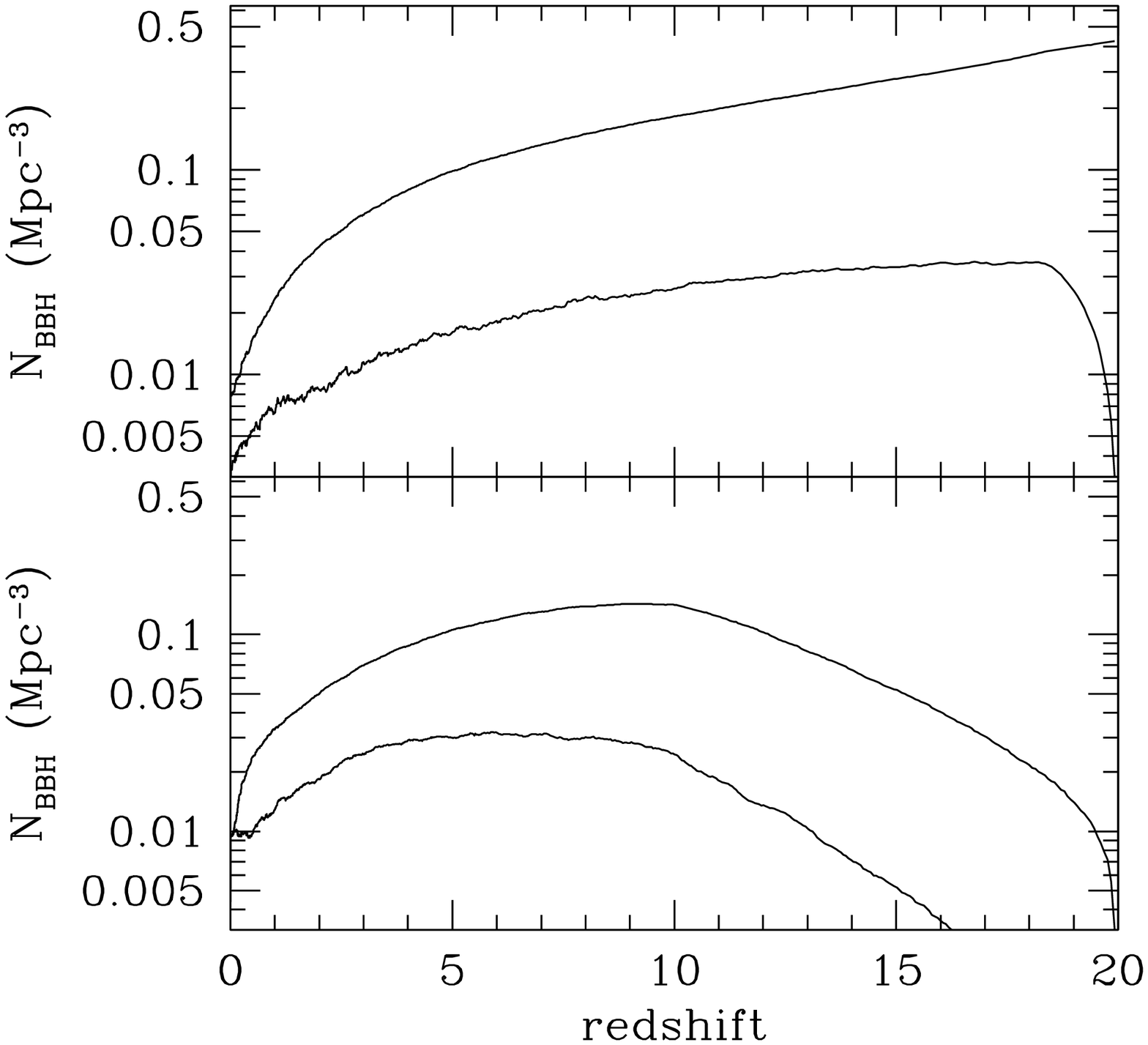,width=\columnwidth}}
\caption{\footnotesize Number density of MBHs ({\it upper curves}) and binaries ({\it lower curves}). The fraction of MBHs in binaries is very small at most times. {\it Upper panel:} direct collapse. {\it Lower panel:} popIII remnants. }
\label{fig2}
\end{figurehere}

Models with large MBH seed masses appear to be favored in a high recoil velocity scenario. However, note that the 1-$\sigma$ dispersions of the two samples, \citep{ferrareseford,tremaineetal2002} are compatible, but not identical. The fraction of outliers can differ up to a factor of two between the two samples for galaxies with velocity dispersion below $200\kms$. The uncertainties and small size of the observational sample contribute to prevent to draw firm conclusions: a larger sample of measured MBH masses, which can be available to future instruments, such as Giant Magellan Telescope, can help assess the relative importance of the gravitational rocket. 

We recall here that we have adopted the orbital configuration yielding the highest recoil velocities, which, as pointed out by \cite{bogdanovic2007} is probably rather uncommon. Also, \cite{campanelli2007b} discuss how the sinusoidal dependence on the orbital phase implies a reduction of $\sqrt 2$ in the root-mean-square recoil velocity, in case of optimal orbital configuration with random phase configuration. The fraction of outliers that we find represents therefore an upper limit, and our results emphasize also the necessity of further investigations on the orbital and phase configuration expected in merging MBH binaries. 

\begin{figurehere}
\centerline{
\psfig{file=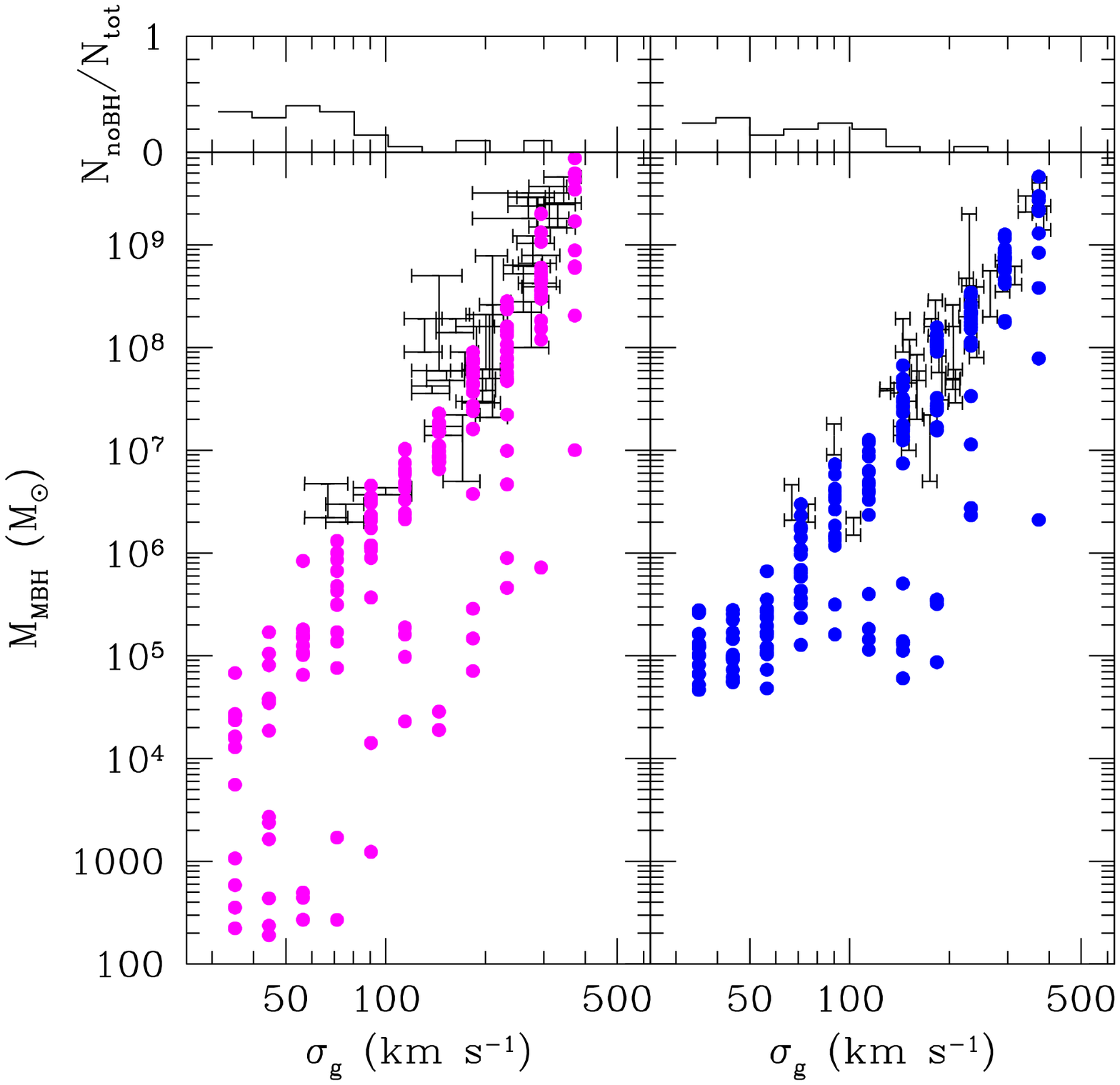,width=\columnwidth}}
\caption{\footnotesize The $M_{\rm SMBH}-$velocity dispersion ($\sigma_c$) relation at $z=0$. Every circle represents the central MBH in a halo of given $\sigma_c$. Observational data are marked by their quoted errorbars, both in $\sigma_c$, and in $M_{\rm SMBH}$ (Tremaine et al. 2002). {\it Left:} popIII remnants; {\it right:} direct collapse. {\it Upper panels:} fraction of galaxies at a given velocity dispersion which do not have a central SMBH.}
\label{fig3}
\end{figurehere}

\subsection{Special cases}
While Figure \ref{fig1} is reassuring with respect to the ``average'' MBH, as the simulations sample a statistical ensemble of galaxy halos, it is not representative of very rare, highly biased regions, where MBH formation is more common, and so are binary MBHs. \cite{Haiman2004} pointed out that the recoil can be indeed threatening the growth of the MBHs that are believed to be powering the luminous quasars at $z\simeq 6$ detected in the Sloan survey \citep[e.g.,][]{Fanetal2001a}.  In fact, in such a biased volume, the density of halos where MBH formation can be efficient (either by direct collapse, or via PopIII stars) is highly enhanced.  
The net result is an higher fraction of binary systems, and binarity is especially common for the central galaxy of the main halo. 
While the ``average'' MBH experiences  at most one merger in its lifetime, a MBH hosted in a rare exceptionally massive halo can experience up to a few tens mergers, and the probability of ejecting the central MBH, halting its growth, is 50-80\% at $z>6$. \cite{VolonteriRees2006} demonstrated that this result is not dependent on the ``optimal'' orbital and phase configuration. This implies that MBHs at high redshift did not mainly grow via mergers, as proposed in works that did not take the gravitational recoil into consideration \citep[e.g.,][]{pelupessy2007,lietal2006}.

Another special case, at the other very end of the MBH mass scale, is represented by merging MBHs in star clusters.  It has been proposed \citep[e.g.,][]{PZ2004,ato2004} that the merging of main-sequence stars via direct physical collisions can enter into a runaway phase, and form a very massive star, which may then collapse to form a MBH. A more exotic outcome of the process can happen in star clusters with a relatively large (but well within the expected values) primordial binary fraction: the formation of two MBHs  \citep{ato2006}.  Star collisions likely transfer angular momentum to the very massive star, and a rotating very massive star plausibly collapses into a spinning MBH \citep{saijo2002}.  If the two star cluster MBHs form with a mass ratio larger than $q=0.1$, when they proceed to coalescence \citep{gultekin2004, gultekin2006,fregeau2006}, the magnitude of the recoil for most configurations  is larger than 50 $\kms$, leading to a high escape probability.  If formation of binary MBHs is common in star clusters, their ejection is highly likely as well.  A similar fate would await MBHs binaries forming following the merger of two star cluster, each with a MBH \citep{pau2006}. 
Although in both cases the MBH binary would be a source of gravitational radiation, which might allow to confirm the predictions of the previous scenario for MBH formation in cluster, after the ejection the MBHs would join the very difficult to detect  population of wandering MBHs  \citep{VolonteriPerna2005,mapelli2006, kuranov2007}.
\citep{future}

\acknowledgements 
We wish to acknowledge enlightening discussions with D. Richstone and K. G{\"u}ltekin.


\end{document}